\documentclass[5p]{elsarticle}

\usepackage{amsmath,amssymb}
\usepackage{hyperref}
\journal{Astronomy \& Computing}

\biboptions{authoryear}

\usepackage{color}

\begin{document}

\begin{frontmatter}

\title{Physical Publicly Verifiable Randomness from Pulsars}


\author[cass,mq]{J. R. Dawson\corref{mycorrespondingauthor}}\cortext[mycorrespondingauthor]{Corresponding author}
\ead{joanne.dawson@csiro.au}
\author[cass]{George Hobbs}
\author[nanjing]{Yansong Gao}
\author[data61]{Seyit Camtepe}
\author[data61,pas]{Josef Pieprzyk}
\author[naoc,ucas]{Yi Feng}
\author[cass,mq]{Luke Tranfa}
\author[cass,qut]{Sarah Bradbury}
\author[naoc,ucas]{Weiwei Zhu}
\author[naoc,ucas,naoc-ukzn]{Di Li}

\address[cass]{CSIRO Space and Astronomy, Australia Telescope National Facility, PO Box 76, Epping, NSW 1710, Australia}
\address[mq]{Department of Physics and Astronomy and MQ Research Centre in Astronomy, Astrophysics and Astrophotonics, Macquarie University, NSW 2109, Australia}
\address[nanjing]{School of Computer Science and Engineering, Nanjing University of Science and Technology, Nanjing 210094, China}
\address[data61]{CSIRO Data61, PO Box 76, Epping, NSW 1710, Australia}
\address[pas]{Institute of Computer Science, Polish Academy of Sciences, Poland}
\address[naoc]{National Astronomical Observatories, Chinese Academy of Sciences, Beijing 100101, People's Republic of China}
\address[ucas]{University of Chinese Academy of Sciences, Beijing 100049, People's Republic of China}
\address[qut]{School of Chemistry and Physics, Queensland University of Technology (QUT), Brisbane, Queensland 4001, Australia}
\address[naoc-ukzn]{NAOC-UKZN Computational Astrophysics Centre, University of KwaZulu-Natal, Durban 4000, South Africa}

\begin{abstract}
We demonstrate how radio pulsars can be used as random number generators. Specifically, we focus on publicly verifiable randomness (PVR), in which the same sequence of trusted and verifiable random numbers is obtained by multiple parties. PVR is a critical building block for many processes and algorithms (including cryptography, scientific trials, electoral audits and international treaties). However, current approaches (based on number theory) may soon become vulnerable to quantum computers, motivating a growing demand for PVR based on natural physical phenomena. In this context, we explore pulsars as a potential physical PVR source. We first show that bit sequences extracted from the measured flux densities of a bright millisecond pulsar can pass standardised tests for randomness. We then quantify three illustrative methods of bit-extraction from pulsar flux density sequences, using simultaneous observations of a second pulsar carried out with the Parkes telescope in Australia and the Five-hundred-metre Aperture Spherical radio Telescope (FAST) in China, supported by numerical simulations. We demonstrate that the same bit sequence can indeed be obtained at both observatories, but the ubiquitous presence of radiometer noise needs to be accounted for when determining the expected bit error rate between two independent sequences. We discuss our results in the context of an imaginary use-case in which two mutually distrusting parties wish to obtain the same random bit sequence, exploring potential methods to mitigate against a malicious participant.
\end{abstract}


\end{frontmatter}

%\linenumbers

\section{Introduction}

Randomness -- defined as the unpredictable outcomes of theoretical or physical processes -- forms the foundation of security, privacy, trust and fairness. Even the most robust cryptographic methods become vulnerable if the underlying randomness is weak or vulnerable. ``Publicly verifiable randomness'' (PVR) emerges as a solution to this concern in multi-party settings. PVR is randomness extracted from a publicly accessible process or object, the output of which can be verified by third parties. A robust PVR protocol should satisfy the following five critical requirements of accessibility and verifiability~\citep{syta2017scalable,schindler2020hydrand,zhangaberand}: \textit{Availability} -- no party is able to block access to the source and each party can access the source at any time \citep{rabin1989verifiable}; \textit{Unpredictability} -- no party is able to predict future random bits; {\it Non-Malleability} -- no party is able to influence future random bits; {\it Public-Verifiability} -- any party is able to verify correctness of generated bits; {\it No-Trusted Server} -- no trusted server is needed to activate and manage the randomness source \citep{syta2017scalable}. 

These requirements can be challenging to fulfil. Current  state-of-the-art  PVR systems are built upon mathematical problems that are believed to be intractable. Many such problems relate to either factoring large integers or finding discrete logarithms \citep{RSA78,DH76}. However, as shown by \cite{Shor97}, both problems are ``easy'' for a quantum computer. 
It is therefore necessary to develop %security primitives without reliance
approaches that do not rely on number theory.

Physical phenomena \citep[e.g.][]{pappu2002physical} are one attractive alternative. Particle diffusion \citep{jiang2017novel}, atmospheric turbulence \citep{marangon2014}, chaos in laser emission \citep{uchida2008fast} and DNA synthesis \citep{meiser2020dna} are just a few examples of physical processes used to generate random sequences. However, even with the rapid progress being made in random number generation from quantum physical processes \citep{pironio2010random, avesani2018, Liu2018}, random number generators based on local natural physical phenomena generally do not fulfil all of the formal PVR requirements. In particular they generally fail non-malleability (a party is able to influence the resulting bits), public-verifiability (it is challenging for other parties to verify the bits) and/or no-trusted server (they may require a single party to provide the random sequences). 

Here we propose the use of pulsars as physical PVR sources. While pulsars are well-known for the extreme precision to which their rotational period can be measured 
\citep[e.g.][]{hobbs+20}, many properties of pulsar emission exhibit apparently random or unpredictable behavior. We may broadly divide these unpredictable characteristics into two categories: (i) those that can be directly measured for each rotation of the pulsar, such as the brightness of each pulse (e.g., \citealt{ritchings76}), or pulse-to-pulse variations in the pulse shape or phase (e.g., \citealt{zhang2019}); (ii) those that are measured over longer time scales, such as tiny month-to-year-scale irregularities in a pulsar's rotational period (e.g., \citealt{hobbs2010}), or glitch, nulling or mode switching events (as described in \citealt{lk12} and \citealt{cordes2013}). 

In this work we focus on the integrated flux densities of individual pulses as our random number generator. While the distribution of intrinsic flux densities from most pulsars are known to follow log-normal or power-law distributions \citep{mickaliger2018}, 
no physical theory exists that can predict a given sequence of individual pulses. That said, no existing work has formally examined the randomness of pulsar flux density sequences (although see \citealt{doyle2011} for an analysis of the ``communicative complexity'' of a pulsar signal), and we therefore perform some initial checks of flux density randomness in this work.  

\begin{figure*}
\centering
\includegraphics[scale=0.6]{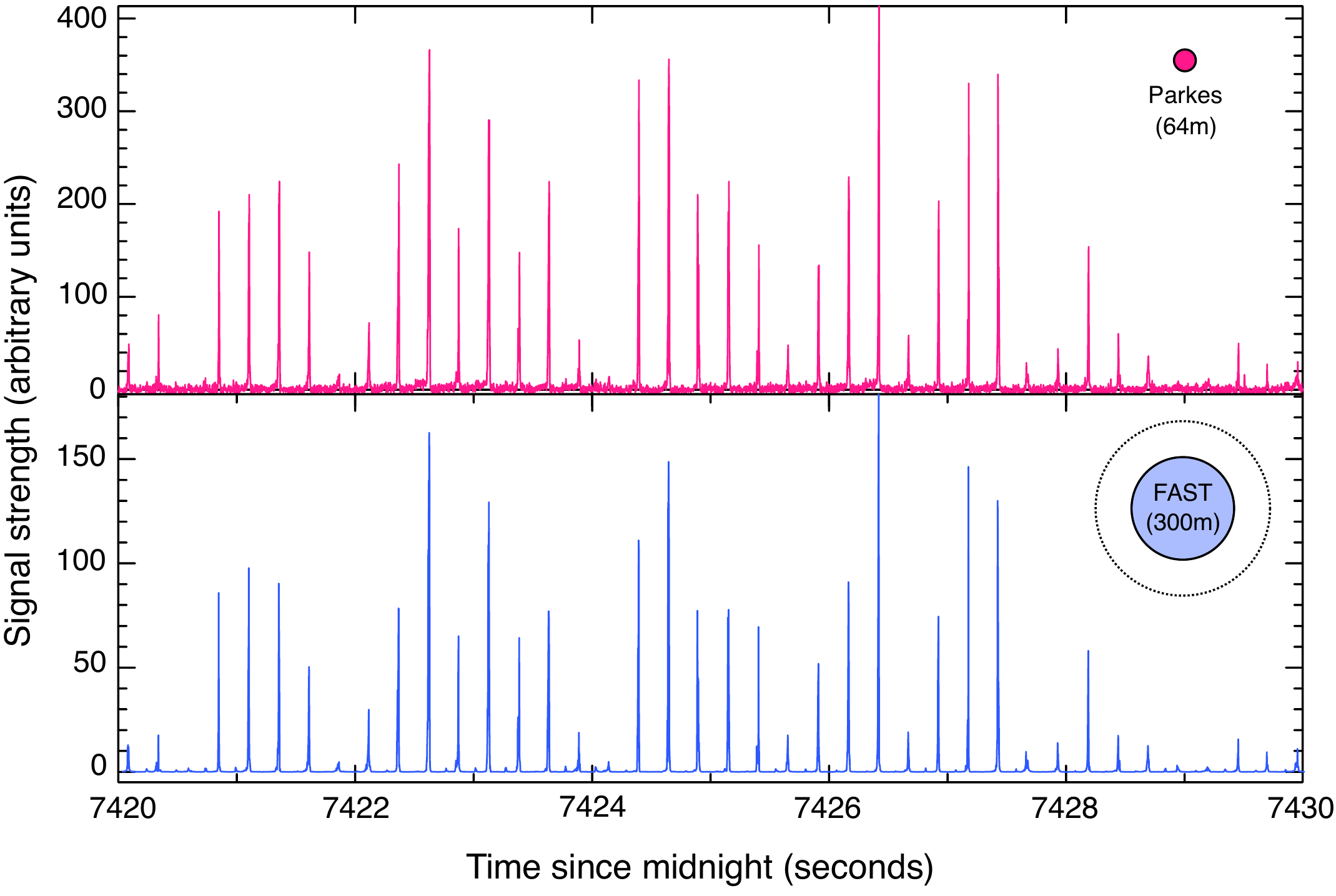}
\caption{Uncalibrated pulse trains for PSR~J0953$+$0755, recorded using the Parkes (upper panel) and FAST (lower panel) on 25/09/2019. The data have been de-dispersed and summed in frequency to produce pulse sequence for each telescope. Circles in the upper right hand corners of the plots represent the diameters of the two telescopes (for the 500m FAST dish only the central 300m is illuminated). Discrepancies in the signal strength at the two facilities are discussed  in the text.}\label{fig:parkesFast}
\end{figure*}

There are a number of examples in the literature of the use of astrophysical sources as randomness generators. The measurables range from CMB power spectra \citep{lee2015}, to the arrival times of cosmic photons \citep{wu2017}, to hot pixels in astronomical images \citep{pimbblet2005} to radiometer noise \citep{chapman2016celestial}. The key distinction of the present work is that we are seeking \textit{publicly verifiable} randomness, in which two or more observers must be able to obtain the same trusted sequence -- i.e. replicability is the desired outcome, not a detriment. 
In this context, pulsars are almost unique. Unlike the examples cited above, multiple parties can independently observe an identical pulse sequence. Unlike observations of sunspots (which can achieve the same multi-party replicability), measurements can be obtained at a rate of many hundreds of pulses per second.

In principle, pulsars can fulfil the key requirements of PVR:
\begin{description}
\item[Availability:]{There are over 3000 known pulsars, meaning that a suitable common source will be available to parties at most geographic locations. (Note that we return to the question of realistic source numbers briefly in Section \ref{sec:outlook}.) The signal cannot be blocked at its source, and pulsars emit over a very broad frequency range, making terrestrial blocking a challenge.} 
\item[Unpredictability:]{Several properties of pulsar emission produce unpredictable time series. (Though see Section \ref{sec:random} for some formal tests of randomness.)}
\item[Non-malleability:]{It is impossible to influence future pulses at their source, and challenging for an adversary to bias the measured signals from Earth.}
\item[Public-verifiability:]{Any participant with a sufficiently sensitive telescope can verify the shared random sequence. (Though we explore in Sections \ref{sec:random} and \ref{sec:discussion} some of the difficulties associated with a practical implementation of that.)}  
\item[No-trusted server:]{The randomness generation does not require a trusted device or trusted server.} 
\end{description}

There are complicating factors of course. These include the spatial scale for scintillation, instrumental effects (differences in observing modes and processing techniques), site-dependent radio frequency interference (RFI), and intrinsic sensitivity differences. In terms of scintillation, the interstellar medium imprints signals onto the pulse sequence, and can brighten or weaken the pulses on time scales from minutes to years (details are provided in \citealt{lk12}). The characteristic spatial scale for scintillation is 1000s of km (e.g., \citealt{narayan92}) meaning that variations in flux density are expected for telescopes that are separated by such distances. Each telescope also adds its local unique Gaussian noise to the observed pulse train, arising primarily from the telescope receiver system (see ~\citealt{hobbs_uwl} and \citealt{li18} for details of the Parkes and FAST observing systems, respectively). While the magnitude of the local noise contribution can be predicted exactly from knowledge of the system, its contribution to any given pulse is random, imposing a telescope-specific layer of additional randomness to the measurements. Instrumental differences and data processing choices may also produce slight differences in the recorded flux densities. All of these effects must be taken into account when attempting to extract a trusted common sequence from multiple telescopes.

In this work we present a first attempt at verifying a common random number sequence from simultaneous observations of the same pulsar. We use observational data from the Parkes 64m dish in Australia and the Five-hundred-metre Aperture Spherical radio telescope (FAST, \citealt{nan11}) in China, first to demonstrate that pulsar flux density sequences can satisfy some common tests for randomness, and then to explore methods of recovering the same sequence of bits from two sets of observations. Section \ref{sec:obs} describes the observations used in our work, including the necessary corrections for differences in observing and processing methods. Section \ref{sec:random} shows that pulsar sequences can pass standarised tests for randomness. In Section \ref{sec:common} we use both observations and numerical simulations to quantify three illustrative methods of bit-extraction from pulsar flux density sequences. In Section \ref{sec:discussion} we qualitatively discuss the implications of our results in the context of an potential real-life use case -- publicly auditable selections between two mutually distrusting parties -- including in the case where there is a malicious participant. We discuss future outlook in Section \ref{sec:outlook}.

\section{Observations}
\label{sec:obs}

We make use of two datasets for this work. The first provides one million pulses from the bright millisecond pulsar, PSR~J0437$-$4715 (with a period of 0.00575s). This source is only observable from Parkes, not FAST, and is used to test the randomness of a long pulse sequence. Simultaneous observations of a second pulsar, PSR~J0953$+$0755 (with a period of 0.253s) are then used to test whether geographically-separated telescopes can be used to obtain a common and trusted bit sequence from an observed pulse train.

PSR~J0437$-$4715 was observed by the Parkes telescope on UTC 03-30-2020 at 01:19:45 with the PDFB4 backend recording the central beam of the 20\,cm multibeam receiver.  We recorded using 512 channels over a 256\,MHz band with a central observing frequency of 1369\,MHz. The polarisation channels were summed and the data quantised to 1\,bit values. The observation lasted two hours, or equivalent $\sim 1.2$\,million pulses from the pulsar.  The \textsc{dspsr} software \citep{ref:dspsr} was used to extract individual pulses, and 5\% of the band edges were removed using \textsc{pazi}.  An analytic template was formed using \textsc{paas} and the flux density values output using \textsc{psrflux} \citep{ref:psrchive}.

For the two-telescope comparison we used simultaneous observations of PSR~J0953+0755 with the FAST and Parkes telescopes.  The original purpose of the data collection was to measure the time delay between the observing systems to support the FAST commissioning (the time delay was determined to be significantly less than 1 pulse period). The observation began at UTC 25-09-2019 at 02:03:11. At Parkes the project identifier was PX500. The PDFB4 backend was used, producing pulsar search-mode data with 2 bit digitisation, at a central observing frequency of 1369\,MHz and a bandwidth of 256\,MHz.  512 frequency channels were produced across the band.  Single pulses were obtained using \textsc{dspsr} and pulsar flux densities determined using \textsc{psrflux}. The FAST produced 4096 frequency channels across a 500\,MHz bandwidth, centred at 1250\,MHz.  8-bit data streams were output.  The data were also processed using \textsc{dspsr} and \textsc{psrflux} (using an analytic template that was independent from the one used at Parkes) to produce the flux density values per pulse.  The resulting data sequences were viewed by eye to confirm that the same pulses were being identified at both telescopes. Note that we explicitly chose not to carry out a careful flagging and calibration process as such processing steps are time consuming and, as described in Section \ref{sec:common}, for the applications considered here we require that the data are streamed in close to real time. 

\begin{figure}
    \centering
    \includegraphics[width=8cm]{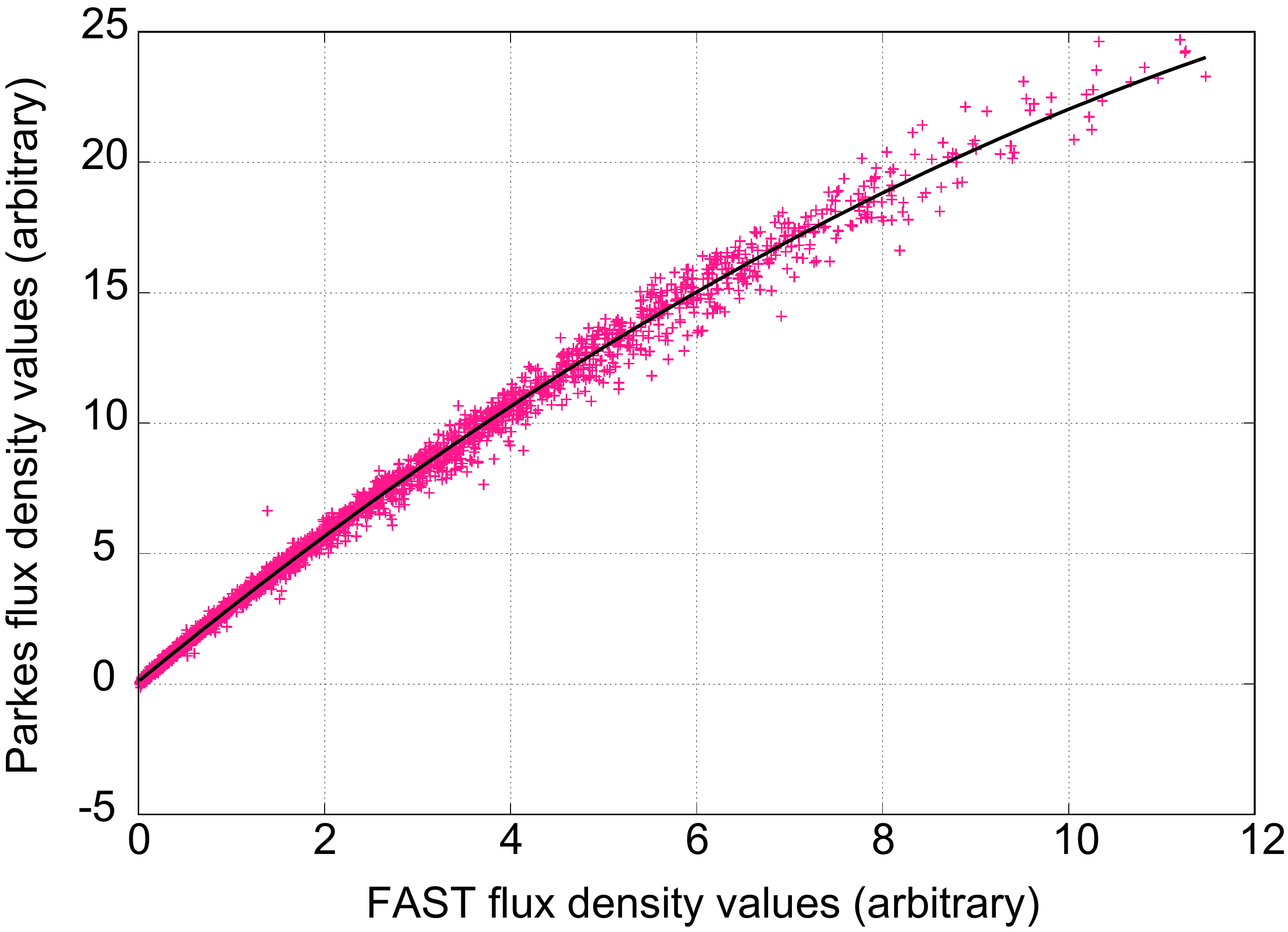}
    \caption{Comparison between the uncalibrated flux density values obtained by through simultanoues observations of PSR~J0953+0755 using the FAST and Parkes telescopes. The line shows the polynomial fit described in the text.}
    \label{fg:compareFlux}
\end{figure}

Figure~\ref{fig:parkesFast} shows a small subset of the pulse sequences from both telescopes. Very broadly, the pulse sequences are similar and the signal to noise is approximately 10 times higher in the FAST data. A more careful look does indicate differences between the data sets. For example the pulse near 7421 seconds since midnight is brighter than the adjacent pulses in the FAST data set, but not for the Parkes data set.

As no calibration steps were carried out, the pulse sequence in Figure \ref{fig:parkesFast} (and hence the flux density measurements) are in arbitrary units. With proper calibration these could be converted to Janskys. However, we would still expect slight variations in the flux density measurements (and their uncertainties) at the two facilities. Firstly, the Parkes and FAST telescopes are separated by $\sim$8000\,km, which is similar to the spatial scale of the scintillation pattern. Secondly, pulsar flux density values and uncertainties are usually measured by cross-correlating an analytic template of the average pulse profile with individual pulses. As individual pulses are known to vary in shape, the analytic template will not be a perfect match, and the corresponding flux density uncertainty will not directly represent the system radiometer noise. 

In our uncalibrated data sets, with 2-bit quantisation for the Parkes telescope output (see e.g., \citealt{vanVleck}), we see systematic differences in both the measured flux density values (Figure~\ref{fg:compareFlux}) and their measured uncertainties. For this reason, along with the scintillation, calibration, and template-matching issues described above, we do not expect a one-to-one relationship between the measured flux densities in our datasets. We therefore identified and applied the quadratic polynomial that is, on average, required to scale the FAST flux density values to the Parkes flux density scale ($f^\prime = 0.1 + 2.928f -0.07f^2$ where $f$ is the FAST flux density value), which is shown as the overlay on Figure~\ref{fg:compareFlux}. The parameter fit is remarkably good and we use this for analysis in the remainder of this paper.

\begin{figure}
    \centering
    \includegraphics[width=8cm]{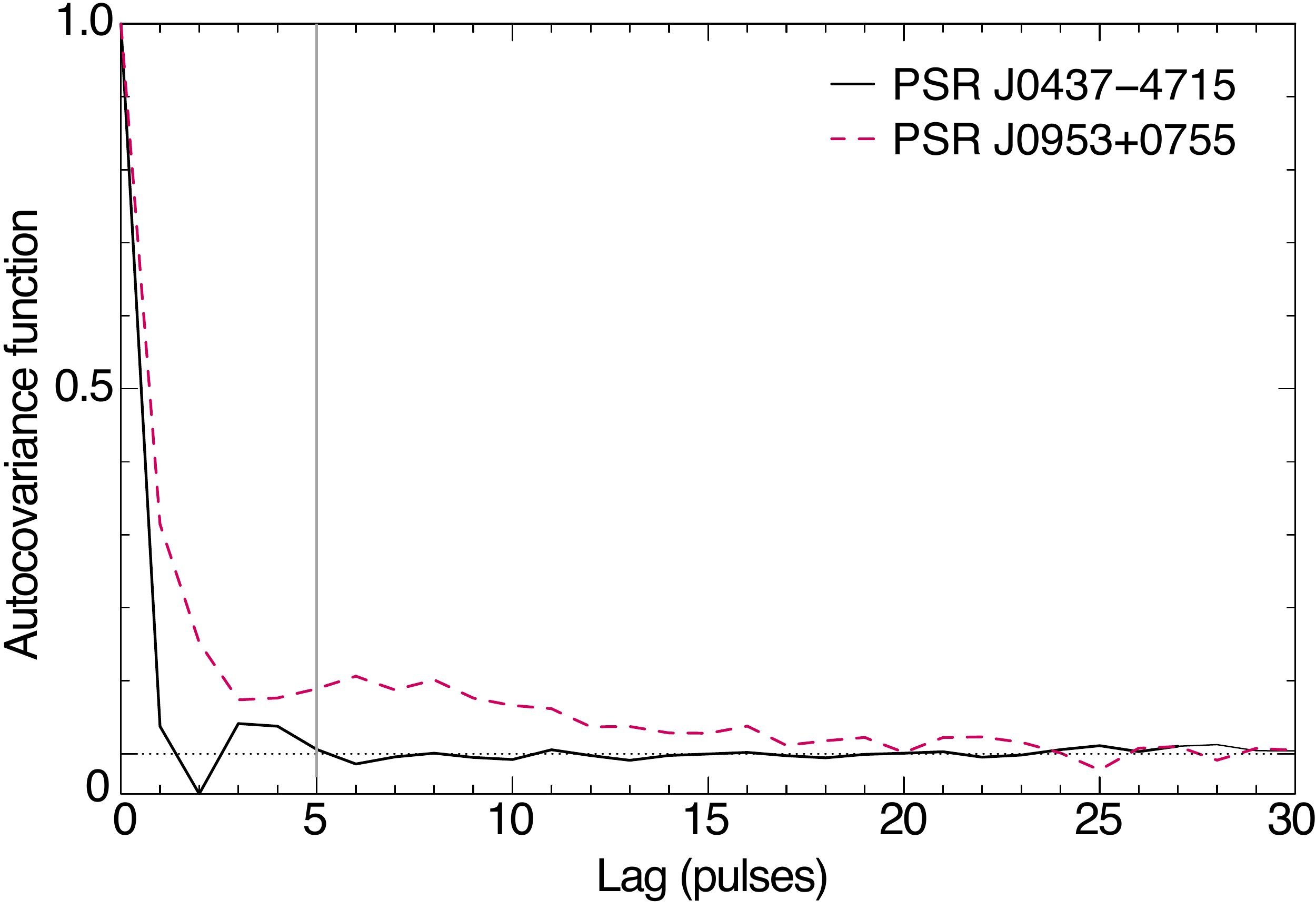}
    \caption{Autocorrelation function for the pulse flux densities for PSR~J0437$-$4715 (black, solid line) and PSR~J0953+0755 (pink, dashed line). Note that PSR~J0437$-$4715 is used for the randomness tests in this work, with the vertical grey line marking the sampling interval chosen to avoid correlations in the extracted sequence. PSR~J0953+0755 is used for the common bit sequence tests. We note that in this testing context the longer correlation timescale is unimportant, but would need to be taken into account in a real life scenario requiring good randomness.}
    \label{fig:acf}
\end{figure}

\section{Analysis and Results}

\begin{table*}[t]
	\centering 
	\caption{Results of randomness tests using the NIST test suite. Details of the tests and bit-extraction methods are described in Sections \ref{sec:random} and \ref{sec:common}. For J0437$-$4715 the number of bits is 200,000 for Method 1 (split into 100 sequences each 2000 bits long), and 240 and 315 for Methods 2 and 3. For J0953+0755, the number of bits is 192. A running median of 500 pulses was used when extracting bits from J0437$-$4715. For Method 2 $N = 100$ and $n_t = 15$. For Method 3 $N=100$ and $f_t = 4$. For all results apart from Method 1 on J0437$-$4715, we test a single sequence and therefore obtain a single P-value, evaluating whether that exceeded the passing threshold of 0.01. For Method 1 on J0437$-$4715, each test produces 100 P-values, and we present the P-value$_T$ results. The  P-value$_T$ for the Discrete FT test is 0.000089 and the P-value histogram is 7,8,7,12,9,18,18,0,17,4 (in bins of 0.1). For 100 binary sequences the minimum pass rate is approximately 96. We use a block length of 32 for the block frequency test and 3 from the approximate entropy test.} .
	\begin{tabular}{lccccccc} 
	\hline \hline
	 & \multicolumn{4}{c}{PSR J0437$-$4715} &&& PSR J0953+0755\\
	 \cline{2-5} \cline{7-8} 
	  & \multicolumn{2}{c}{Method 1} & Method 2 & Method 3 &&&  Method 1 \\ 
	 NIST test & Passes (Pass) & P-value$_T$ (Pass) & P-value (Pass) & P-value (Pass) &&& P-value (Pass) \\ \hline
	 Frequency  & 99/100 (Y) & 0.28 (Y) & 0.61 (Y)  & 0.40 (Y) &&& 0.89 (Y) \\
	 Block Frequency & 99/100 (Y) & 0.21 (Y) & 0.43 (Y) & 0.12 (Y) &&& 0.62 (Y) \\
	 Cumulative Sums  & 98/100 (Y) & 0.02 (Y)& 0.78 (Y) & 0.29 (Y) &&& 0.39 (Y) \\	
	 Runs            & 99/100  (Y)& 0.51 (Y) & 0.36 (Y) & 0.73 (Y) &&& 0.39 (Y) \\
	 Longest Runs of 1s  & 99/100 (Y) & 0.67 (Y) & 0.90 (Y) & 0.82 (Y) &&& 0.77 (Y) \\		
	 Discrete FT  & 98/100 (Y) & 0.00 (N) & 1.00 (Y) & 0.40 (Y) &&&  0.89 (Y) \\	
	 Approx. Entropy & 98/100 (Y) & 0.38 (Y) & 0.51 (Y) & 0.89 (Y) &&& 0.02 (Y)\\ \hline\hline
	\end{tabular}
	\label{tab:newnist} 
\end{table*}

The questions we aim to address are: 1. Can pulsars provide sequences that satisfy common tests for randomness? 2. Using two telescopes observing the same pulsar, is it possible to obtain a common bit sequence that both parties can trust? Randomness tests are most meaningful when performed on long sequences. Common bit sequence extraction requires that we have simultaneous observations of the same source. In this work, we perform randomness tests primarily on our one million pulse dataset from PSR~J0437$-$4715, with only minimal testing possible on the shorter pulse sequence from PSR J0953+0755. While all pulsars are unique in their emission characteristics, tests on a single source provide an important proof-of-concept. We then explore bit sequence extraction on 3844 pulses from PSR J0953+0755, observed simultaneously by Parkes and FAST.

\subsection{Testing for Randomness}
\label{sec:random}

We utilise the NIST SP800-22b test suite \citep[hereafter NIST,][]{NIST2010} -- a commonly used collection of standardised tests for randomness -- to test whether bit streams extracted from the sequence of pulse flux densities of PSR~J0437$-$4715 satisfies basic randomness tests. We also perform a single test on the much shorter sequence from PSR J0953+0755. An autocovariance function of the pulse flux density for the two sources (see Figure~\ref{fig:acf}) shows that neighboring pulses are correlated, indicating flux density variations on timescales longer than the pulsar rotational period. 
Such behaviour is not unexpected: individual pulses are known to show complicated structure and pulse-to-pulse correlations \citep[see e.g.][]{zhang2019}, the origins of which are not well understood. For PSR~J0437$-$4715 this covariance drops to almost zero for separations of five pulses or more. We therefore retain only every 5th value, for a total pulse sample of 200,000 pulses. For PSR J0953+0755 the timescale is much longer, around 20 pulses. Retaining only every 20th value leaves us with 192 pulses for this source.

The NIST suite tests against the null hypothesis of a uniform, random distribution of binary bits. We extract suitable bit sequences in three ways, corresponding to the three methods of bit-extraction outlined in Section \ref{sec:common}. For Method 1 (simple median), pulses are set to 1 if they fall above the median for the dataset and 0 if they fall below. For PSR~J0437$-$4715 we must apply a running median, to remove variations due to instrumental effects and scintillation over the 100 minutes of observations. The median window is set to 500 pulses, corresponding to $\sim$3\,sec (we also tested window sizes between 50 and 5000 with no effect on the pass rates). All other Methods are as described in \ref{sec:common}, with details on thresholds given in the caption to table \ref{tab:newnist}. We note that randomness tests using Methods 2 and 3 can only be performed on PSR~J0437$-$4715, since the total number of output bits for PSR J0953+0755 ($\sim$10) does not meet minimum requirements for bit stream length.

We then consider only the seven tests for which meaningful results can be obtained from our sample sizes. Our longest sequence is 200,000 bits (PSR~J0437$-$4715, Method 1). For this sequence we analyse 100 bit-streams each 2,000 bits in length, using the default NIST suite parameters. All other bit-streams range from 192 to 315 bits in length, permitting only one iteration of each test. Full details of the tests and their statistical analysis are presented in \cite{NIST2010}, but in brief, they are:
\begin{itemize}
    \item {\bf Frequency:} tests the proportion of 0s and 1s in the sequence.\vspace{-0.2cm}
    \item {\bf Block Frequency:} as above, but for smaller segments of the entire sequence.\vspace{-0.2cm}
    \item {\bf Cumulative Sums:} the cumulative sum of the bits should follow a random walk. The properties of the random walk are probed in this test.\vspace{-0.2cm}
    \item {\bf Runs:} identifies the number of adjacent identical bits (either 0 or 1).  \vspace{-0.2cm}
    \item {\bf Longest Runs of Ones:} as above, but identifies the longest run of 1s within small segments of the data.\vspace{-0.2cm}
    \item {\bf Discrete Fourier Transform:} searches for repetitive patterns in the sequence.\vspace{-0.2cm}
    \item {\bf Approximate Entropy:} determines the frequency of overlapping short patterns in the entire sequence.
\end{itemize}

For each sequence, the NIST tests provide P-values representing the probability that the randomness of the sequence is equivalent to that produced by a perfect random number generator. The P-value must be bigger than the Type-I error (i.e., true hypothesis rejected) probability to accept the hypothesis that the sequence is random. For multiple blocks extracted from the same long sequence, these P-values 
are subsequently used to analyse (1) the proportion of sequences that pass the specified test and (2) the uniformity of the P-value distribution (described by P-value$_T$). A minimum pass proportion and P-value$_T$ are then required for a sequence to pass this more robust two-level testing. In our case, two-level tests can only be carried out for the 200,000 pulse sequence from PSR~J0437$-$4715 using Method 1.

The results are shown in Table~\ref{tab:newnist}. All sequences pass all of the tests, with the exception of the Discrete Fourier Transform test, for which the 200,000-bit pulse sequence from PSR~J0437$-$4715 fails the final P-value uniformity check. 
The reason for this is unclear, but we note that the standard NIST implementation of this test has been shown to produce erroneously low P-value$_T$ for known good RNGs, particularly in the case of long sequences \citep[e.g.][]{kim2004, pareschi2012}.

These simple checks should be considered a preliminary step in a more thorough exploration of pulsar randomness, but we consider them sufficient evidence to proceed under the assumption that pulsar flux density sequences can provide reasonable random number sequences. 
Future analysis might include a broader suite of statistical tests \citep[e.g.][]{wang2015,lorek2020} and (of course) tests on multiple sources. It is also worth noting that physical randomness sources do in general tend to be slightly biased and non-uniform \citep{avesani2018, Liu2018}, and that common post-processing steps (e.g. hashing) are often performed to ameliorate non-randomness, if appropriate for a given application (e.g. \citealt{kwok2011comparison}).

\subsection{Extracting a Common Number Sequence from Multi-telescope Observations}
\label{sec:common}

\begin{figure*}[t]
	\centering
\includegraphics[width=16cm]{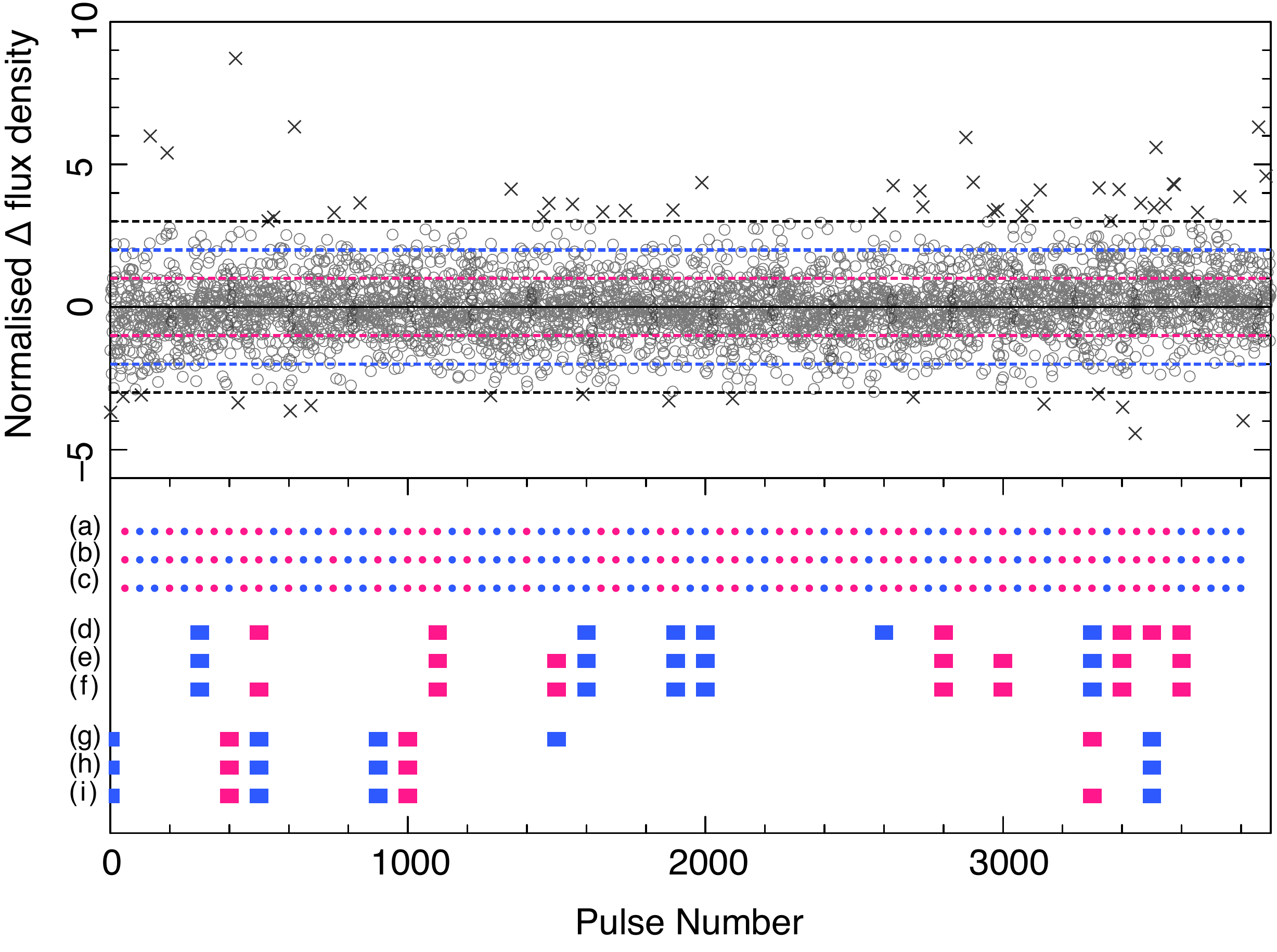}
	\caption{Results of verification and bit-extraction processes for the Parkes and FAST observations of PSR~J0953+0755. The upper panel shows the normalised flux density difference between measurements at the two observatories with horizontal dashed lines indicating 1,2 and 3$\sigma$. The grey crosses indicate pulse numbers in which the two telescope results deviate by more than 3$\sigma$, rejected in further analysis. The bottom panel shows the three bit extraction methods described in the text, where pink corresponds to 1 and blue to 0. Labels (a), (b) and (c) correspond to the bits obtained using Method 1 on the Parkes, FAST and averaged data streams, respectively. Labels (d), (e) and (f) correspond to the same three items for Method 2, and (g), (h) and (i) to Method 3. For clarity we only show every 50th bit for Method 1.}
	\label{fig:compareMethods}
\end{figure*}

We now use the simultaneous Parkes and FAST observations of the second pulsar, PSR J0953+0755, together with numerical simulations, to quantify some approaches to extracting common bit sequences from pulsar flux densities. 

In the following we will explore three methods of bit extraction, all of which take blocks of pulse flux densities of length $N$, from which a bit will be extracted if a condition is fulfilled. We define $N_\mathrm{bit,1}$ as the number of bits extracted from telescope 1 and $N_\mathrm{bit,2}$ as the number of bits extracted from telescope 2.  $N_\mathrm{bit,com}$ is the number of blocks for which a bit was obtained from both telescopes. We define a bit error rate (BER) as the fraction of $N_\mathrm{bit,com}$ with mismatching bits. We also perform all methods on a combined flux density sequence made from a simple mean of the datasets from the two telescopes. The number of bits obtained from this is $N_\mathrm{bit,com,av}$.

Pulse flux density differences normalised by the measurement uncertainties should follow a normal distribution. The normalised difference measurements for the Parkes and FAST data are shown in the upper panel of Figure~\ref{fig:compareMethods}. The majority of the flux density values are consistent to within the expected uncertainties, but a handful deviate more as a result of local RFI. For the moment we simply remove all observations that deviate by more than 3$\sigma$, but consider below whether this approach could be exploited by a malicious participant to bias the results. After this pre-selection, we have retained 3844 flux density values from each data stream. 

We also generate a simulated data set of 1 million pulses drawn from a log-normal flux density distribution.  We initially select the mean, $\mu$, and standard deviation, $\sigma$ of the log-normal distribution to be similar to the measurements of PSR~J0953$+$0755, which gives $\mu = 0.98$ and $\sigma = 1.1$.  As pulsars exhibit a range of single pulse flux-density distributions we also trial $\mu = 0.5$ and $\sigma = 0.8$.  For each set of parameters we generate two time series to represent the output from two different telescopes, where each telescope system adds its own Gaussian noise ($\epsilon_1$ and $\epsilon_2$ respectively). For the first simulation we set $\epsilon_1 = 0.06$ and $\epsilon_2 = 0.003$, as a rough approximation of the different noise levels in the Parkes and FAST datasets.   For the second we assume the white noise levels are closer and higher and set $\epsilon_1 = 0.3$ and $\epsilon_2 = 0.2$. The results of all tests for both the real and simulated data are described below, and tabulated in Table \ref{tab:Tests}.

\textbf{Method 1: Simple Median.} We determine a median from the full 18-minutes of observations, and the full one million simulated pulses, and extract one bit per pulse, where a 1 represents a flux density value above the median and 0 below. Out of the 3844 pulses in the real data, this procedure leads to 42 mismatches and hence a bit error rate of $\sim 1\%$. Figure~\ref{fig:compareMethods} shows the output of this method, 
with the bits obtained from the Parkes data labelled (a), the FAST data labelled (b) and the bits obtained from the averaged input sequence as (c). Note that only every 50th bit is plotted for clarity.  In the simulation with input parameters similar to the actual observational data we obtain a similar BER of 0.7\%.  However, we note that the BER for this method is highly dependent on the input parameters for the radiometer noise levels and the pulse properties.  For the second simulation we obtain a much higher BER of 8.6\% with this method.

\begin{table*}[]
    \centering
        \caption{Results of the three bit-extraction methods detailed in the text, as applied to 3844 common pulses from PSR J0953$+$0755 (top subtable) and the one million simulated pulses (bottom subtable). Subscript 1 refers to the telescope with the lower noise (FAST in the real dataset) and subscript 2 to that with the higher noise (Parkes in the real dataset). N$_{\rm bit,1}$ is the number of bits recorded by telescope 1, N$_{\rm bit,2}$ the number recorded by telescope 2, N$_{\rm bit,com}$ the number of blocks for which a bit was recorded at both telescopes (common bits), and BER is the bit error rate for the common bits. N$_{\rm bit,com,av}$ gives the number of bits extracted when the flux density datastreams are first averaged together using a simple mean. All parameters are as described in the main text.}
    \label{tab:Tests}
    \begin{tabular}{l|l|lllll|l}
    \hline \hline 
         \multicolumn{8}{c}{\textbf{Observational data} (PSR J0953+0755)}\\
         \hline \hline 
         {\bf Method} & {\bf Parameters} & {\bf N$_{\rm bit,1}$} & {\bf N$_{\rm bit,2}$} & {\bf N$_{\rm bit,com}$} & \textbf{N$_{\rm bit,2}$/N$_{\rm bit,com}$} & {\bf BER} &  {\bf N$_{\rm bit,com,av}$} \\  \hline
         & & & & & & & \\
         1. Simple Median& $N=1$ & 3844 & 3844 & 3844 & 1.0 & 1.1\% & 3844 \\ 
         & & & & & & & \\
         2. Block Median& $N = 100$, $n_t = 15$ & 11 & 12 & 9 & 1.3  & 0.0\% & 12 \\
         & & & & & & & \\
         3. Difference& $N = 100$, $f_t = 20$ & 6 & 8 & 6 & 1.3 & 0.0\% & 7 \\ 
         & & & & & & & \\
         \hline \hline
         \multicolumn{8}{c}{\textbf{Simulations} (\bf $\epsilon_1 = 0.06$, $\epsilon_2 = 0.003$)}\\
         \hline 
         & & & & & & & \\
          1. Simple Median& $N=1$ & $10^6$ & $10^6$ & $10^6$ & 1.00 & 0.7\% & $10^6$ \\
          & & & & & & & \\
        2. Block Median& $N = 100$, $n_t = 15$ & 1339 & 1329 & 1199 &  1.11 & 0.0\% & 1341 \\
        & & & & & & & \\
         3. Difference& $N = 100$, $f_t = 20$ & 9290 & 9287 &  9283 & 1.00 & 0.2\% & 9288 \\
         & & & & & & & \\
         \hline 
        \multicolumn{8}{c}{\textbf{Simulations} (\bf $\epsilon_1 = 0.3$, $\epsilon_2 = 0.2$)}\\
        \hline
          & & & & & & & \\
         1. Simple Median& $N=1$ & $10^6$ & $10^6$ & $10^6$ & 1.00 & 8.6\%  & $10^6$ \\ 
& & & & & & & \\
        2. Block Median & $N = 100$, $n_t = 15$ & 1266 & 1300 & 725 & 1.79 & 0.0\% & 1324 \\
        & $N = 100$, $n_t = 3$ & 7578 & 7603 & 6151 & 1.24 & 6.0\% & 7576 \\
          & $N = 100$, $n_t = 5$ & 6036 & 6142 & 4524 & 1.36 & 1.8\% & 6083 \\
         & $N = 100$, $n_t = 10$ & 3604 & 3667 & 2398 & 1.53 & 0.2\%  & 3574 \\ 
         & & & & & & & \\
         3. Difference & $N = 100$, $f_t = 20$ & 703 & 703 & 673 & 1.04 & 0.0\% & 699 \\
         & $N = 100$, $f_t = 5$ & 9962 & 9963 &  9947 & 1.00 & 4.4\%  & 9964 \\
         & $N = 100$, $f_t = 10$ & 5727 & 5725 &  5504 & 1.04 &  1.4\% & 5709 \\
         & $N = 100$, $f_t = 15$ & 1937 & 1933 &  1856 & 1.04 &  0.3\%  & 1934\\ 
         & & & & & & & \\
         
        \hline \hline 
    \end{tabular}
\end{table*}

\textbf{Method 2: Thresholded block median using redundant information.}  

In cases where we have many more pulses than required bits, we may make use of the redundant information to reduce the bit error rate. We divide the full sequence of pulse flux densities into multiple blocks of length $N$, and for each block determine the number of flux density values above ($n_\uparrow$) and below ($n_\downarrow$) a  median value that is obtained on a longer time-scale than the block size, but shorter than expected variations caused by gain changes or scintillation. 
We then choose a threshold value, $n_t$, such that a block is only used if $(n_\uparrow - n_\downarrow) > n_t$ or $(n_\downarrow - n_\uparrow) > n_t$. For retained blocks, we assign a value of 1 for $(n_\uparrow - n_\downarrow) > n_t$ and 0 for $(n_\downarrow - n_\uparrow) > n_t$. 
The threshold and block length may then be tuned to generate the required number of bits. For example, dividing the data into blocks of 100 pulses with $n_t = 15$ produces 9 bits that are identical between the data sets, corresponding to a BER of zero. However, each telescope dataset does provide extra (non-common) bits. 
If the flux density sequences are first averaged, we obtain 12 common bits. These results are graphically shown in the lower panel for Figure~\ref{fig:compareMethods}, with the bits obtained using Parkes labelled (d), using FAST labelled (e) and using the averaged input sequence as (f). With these choices for $N$ and $n_t$ both simulations also have BER = 0\%, but each data stream also generates non-common bits at a rate determined by the simulated radiometer noise. Some other results from illustrative parameter values for the simulated data are shown in Table \ref{tab:Tests} for comparison. Tuning the parameters of this method allows us to make the bit error rate arbitrarily low. As we analyse blocks of 100 pulses at a time we are also not affected by pulse-to-pulse correlations. However, unless we first combine the two data streams, there is not a one-to-one mapping of the extracted bits, i.e. $N_\mathrm{bit,1}/N_\mathrm{bit,com}$ and $N_\mathrm{bit,2}/N_\mathrm{bit,com} > 1$. %Questions remain on how best to deal with these non-common bits.

\textbf{Method 3: Difference between adjacent pulses using redundant information.}
We also test the use of adjacent pulses where the difference in flux density exceeds a threshold, $f_t$, which is chosen to be significantly greater than the measurement uncertainty. Again dividing the sequence into blocks of $N = 100$ pulses, we identify in each block the pair of adjacent pulses with the largest deviation, retaining it only if it exceeds $f_t$. The bit is then set to 1 if the pulse pair are in the second half of the block, or 0 otherwise. For $f_t = 20$ (corresponding to $\sim$80\,sigma) we obtain 8 output bits from Parkes, labelled as (g) in the lower panel of Figure~\ref{fig:compareMethods}, and 6 output bits from FAST, labelled (h). (i) shows the common bit-sequence obtained if the flux density sequences are first averaged. The 6 common bits obtained by this method have BER = 0\%. The same threshold tested in the million-pulse simulations produces similarly low BERs ($\sim0\%$), and generates far fewer non-common bits than Method 2. As in Method 2, the BER can be traded off against the number of output bits in the final sequence, and we show in the Table some typical values obtained from the simulations. Again, unless we first combine the two data streams, 
$N_\mathrm{bit,1}/N_\mathrm{bit,com}$ and $N_\mathrm{bit,2}/N_\mathrm{bit,com} > 1$. We note that this method makes use of adjacent pulses and so is not affected by longer-time-scale scintillation effects.

\section{Discussion}
\label{sec:discussion}

Even in an ideal case where we can ignore scintillation and differences in processing and instrumentation, the presence of radiometer noise means that it is never possible to recover the same infinitely long sequence from two telescopes with 100\% fidelity (unless the data streams are first combined). Our results demonstrate this in practice. What constitutes an acceptable error rate in a real-world situation then depends on the desired application and level of participant trust. If the participants share, compare and average their raw flux density sequences prior to bit extraction, a common sequence can always be obtained, but questions surrounding trust and security in a real-world application are not eliminated, just changed. Although quantifying the behavior of various bit-extraction methods is a purely mathematical exercise, meaningful discussion requires us to consider the context of their use in real-world applications. 

We will now consider an imagined use-case in which two mutually distrusting parties wish to obtain the same one-time random number sequence for a specific purpose, such as a publicly-auditable selection. This is a realistic application: for instance, Cordes (private communication) investigated the use of pulsar observations for generating a mutually-verified random number sequence to aid in enforcing Soviet and United States strategic arms limitation agreements in the late 1970s (although this work was not further developed or published).

There are a large number of practical options when designing such a procedure, ranging from the selection of the target source, to pre-processing choices, to data streaming procedures and encryption. For the purposes of discussion we will assume (in common with our test dataset) that the incoming data stream is de-dispersed at each observatory, and uncalibrated flux density values and uncertainties for each pulse transmitted (openly and unencrypted), with a delay less than the pulse period (0.25 seconds for PSR J0953$+$0755). This streaming of the results in close to real time precludes the use of time-consuming pre-processing algorithms, or manual flagging of data affected by RFI, but is an important layer of protection against a bad-faith actor (as we discuss further below). 
Mimicking the data-scaling and outlier removal carried out in Sections \ref{sec:obs} and \ref{sec:common}, we assume that each party first statistically compares the received flux density values with their own sequence, and carries out a scaling procedure to map the measurements to a common scale. The two parties then agree to remove all observations that deviate by more than 3 sigma, and perform bit extraction either on their own data streams individually, or on an average of both streams, using one of the bit-extraction methods explored above. In the former case, they agree to remove any discrepant bits from consideration and retain only the common bits for their final sequence.

Unfortunately, while this protocol is a promising starting point, it is not fully secure against a malicious participant who wishes to influence the results. In common with the cryptography literature, we will refer to this bad-faith actor as ``Chuck''\footnote{The cryptography literature commonly uses Bob and Alice as the primary characters, Eve as an eavesdropper and Chuck as a malicious participant.}. 
Chuck has many options at his disposal if he wishes to corrupt the randomness of the output. For example, any discrepant flux density that is significantly larger than the measurement uncertainty will be identified and removed in the initial verification stage. This provides Chuck with a method of removing certain pulses from consideration prior to bit extraction, by introducing ``fake RFI'' into his own data stream to ensure a pulse is rejected. One possible way to guard against this is to introduce multiple monitors -- i.e. neutral third-parties with their own observatories. This allows us to establish a protocol where the non-affected pulses from other observatories are retained (provided they agree), and discrepant values at a single observatory do not affect the output sequence. 

Alternatively, Chuck may make small-scale changes to his flux density values in an attempt to evade detection. In Method 1, for example, Chuck need only nudge the flux density values for specific bits to one side or the other of the local median in order to bias the data stream. Provided the change remains statistically small, it will be undetected. A straightforward defense against any such behaviour (and indeed the previous example of forcing the removal of a pulse) is to pass the final, agreed-upon bit sequence through a one-way mathematical function (e.g. a hash function). The benefit of this is two-fold. Firstly, because the flux density sequence is streamed live, Chuck cannot know the impact that changing and committing the current bit will have on the hash output, due to the unpredictability of future flux density values. Secondly, Chuck cannot know which bits to change to obtain the desired hash output due to the hardness of the underlying mathematical problem. Chuck can still make arbitrary changes, but cannot predict their result. 

If we wish to avoid reliance on one-way mathematical functions and use only the intrinsic randomness of the pulsar itself, the design of the bit-extraction and verification protocol is our primary means of defense. For example, in Method 2, Chuck must modify a larger number of bits than Method 1, increasing the risk of detection. The real-time streaming of data also mitigates against deliberate bias because Chuck does not know ahead of time (a) what the overall median of the entire data set will be, and (b) whether a given block of 100 is likely to trigger the $n_t$ threshold, until a large fraction of those 100 pulses have already been committed. We note that in the case of millisecond pulsars the required time between detection and transmission would need to be $\lesssim0.5$s in this 100-pulse case. This is feasible: numerous real-time pulse detection pipelines are now in common use at observatories \citep[in particular for searching for fast radio bursts, e.g.][]{agarwal2020}. However, if there is a delay in the ability to transmit the data sequence then the block length could be increased as required. Furthermore, since the theoretical BER for Method 2 can be arbitrarily low, Chuck  likely cannot \textit{flip} a bit without detection. Chuck's options are then limited to switching his bit off. Here the multi-telescope case becomes useful again. Even assuming the random distribution of flux densities in the desired group of 100 allowed Chuck to switch off a bit without raising suspicion, the verification protocol would have to be such that a non-bit from a single observatory causes the common bit from the others to be rejected. 

Looking now at Method 3, to trigger the $f_t$ threshold for bit-extraction, both parties must measure a single very bright pulse immediately adjacent to a weak one. Faking a bright pulse to ensure a bit is counted is unlikely to help Chuck, since this will likely not be replicated at the partner observatory. He could, however, nudge marginal pulses up slightly in order to increase the chances of both parties triggering the threshold for any given pulse pair. He could also ensure that a bit is not triggered by reporting the presence of RFI to corrupt strategic pulses, although for long pulse trains this could arouse suspicion if applied too selectively. As for the other methods, the most effective mitigation strategy is likely to be multi-party verification and/or hashing to further randomise the outcome of any tampering.

\section{Outlook}
\label{sec:outlook}

In this work, we have attempted to motivate the use of pulsars as a source of publicly-verifiable randomness. As a physical PVR source, pulsars do not suffer from the security threat posed by quantum computing, which can tackle the supposedly intractable mathematical problems upon which current state-of-the-art PVR systems rely. As distant astrophysical objects observable by parties at well-separated geographic locations, they provide randomness without a central trusted server. Producing up to 100s of pulses per second, they emit pulse sequences with random characteristics that can be communicated and verified by geographically dispersed participants. While it is clear that more development is needed, particularly surrounding secure protocols, it may be that these concerns can be easily mitigated if we are happy to use a pulsar-generated number sequence as a real-time random seed for a one-way hash function, rather than as-is.

Source selection is an important consideration. The majority of the $\sim3000$ known radio pulsars are not suitable for single-pulse-based PVR of the kind discussed in this work. Many are so weak that even FAST (currently the world's largest telescope) is unable to detect their individual pulses. In addition, multiple telescopes must be able to observe the same pulsar simultaneously. Sources that are too far North can never be observed in the Southern Hemisphere, and vice versa. Similarly, it is impossible to commensally observe any source from completely opposite sides of the Earth (e.g. from parts of Australia and Europe).

For the Parkes observatory, we estimate that an approximate flux density threshold of $\sim$10 mJy provides sufficient single pulse sensitivity for the algorithms explored in this work\footnote{assuming a system temperature of 20 K, a telescope gain of 0.7~K/Jy, and an observing bandwidth of 256~MHz}. The ATNF pulsar catalogue \citep{manchester2005} lists 63 pulsars above this threshold at 20~cm, of which 47 are observable from the Parkes observatory. Assuming FAST is 10 times more sensitive (a conservative estimate), this number increases to 524 pulsars, about half of which are observable. Clearly this kind of discussion can be generalised to any number of observatories. Even relatively small, low-sensitivity radio telescopes can observe individual pulses from a handful of sources (of the order of 10 to 50 common sources might be expected for typical observatories), and while a pulsar PVR network might not necessarily be fully global in reach, it could span nations and continents. It is also worth noting that a deep-space-based system would suffer from no geographic limitations. 

We have shown that even a single observable property, the pulse flux density, provides numerous options for random number extraction. Future work might consider alternative measurables such as arrival time jitter, long-term timing noise, the time between glitch, null or giant pulse events, the null duration, pulse shape properties and many more (as described in \citealt{lk12}). As one of the primary requirements of PVR is non-malleability, the persistence of random phenomena over wide bandwidths is important (since it mitigates against terrestrial interference attempts). 
Ideally we would demonstrate that a common random sequence could be obtained even in completely different observing bands. To date, all evidence suggests that single pulse variability is broadband over commonly used radio bands, e.g. work such as \citet{moffett97} has shown that the giant pulses from the Crab have an emission bandwidth of at least 3.5\,GHz at radio frequencies. However, an in-depth study with a large sample of pulsars has not yet been performed.

Pulsars are not the only astronomical source that could potentially provide PVR (mention has been made of using sun spots, \citealt[e.g.][]{canetti2007}, or segments of images of the moon), but they are the only known examples that emit over a broad wavelength range (making them difficult to jam), can operate through the day and night and can provide relatively high bit-rates. 

Here we explored a first proof-of-concept -- a single well-defined use case in which two mutually distrusting participants wished to obtain a short random bit sequence. But there may be wider applications for a robust, multi-telescope PVR system, provided that appropriate protocols can be developed. Numerous applications require access to PVR, from lottery games to new blockchain proposals, to protocols for anonymous browsing (including Tor anonymity services), financial audits, electronic voting protocols, international treaties, and publicly-auditable selections \citep{schindler2020hydrand}. It is relatively straightforward to imagine, for example, the selection of national lottery numbers based on Parkes observations, verified by a central regulator with their own telescope(s). 
It is also worth noting that although we have chosen to focus on \textit{public} randomness, future work might consider whether pulsars could be employed more broadly in a range of cryptographic applications. A source of true unbiasable randomness is a crucial building block for cryptographic services, and cryptographic protocols have been broken because of the failure of pseudorandom bit generators. Astrophysical randomness sources may provide a future solution to these problems. 

\section*{Acknowledgements}
We thank the anonymous reviewer for their feedback which helped to improve this manuscript. This work was partially supported by the National Natural Science Foundation of China (grant nos. 11988101, 11725313).
The Parkes radio telescope (Murriyang) is part of the Australia Telescope National Facility, which is funded by the Commonwealth Government for operation as a National Facility managed by CSIRO. FAST is a Chinese mega-science facility, operated by the National Astronomical Observatories, Chinese Academy of Sciences. This paper includes archived data obtained through the CSIRO Data Access Portal (data.csiro.au). LT acknowledges a scholarship from CSIRO Data61. WZ is spported by the National Natural Science Foundation of China 12041303, and 11873067, the CAS-MPG LEGACY project and the National SKA Program of China No. 2020SKA0120200.  JP has been supported by Australian Research Council (ARC) grant DP180102199 
and Polish National Science Center (NCN) grant 2018/31/B/ST6/03003.

\bibliography{Random.bib}

\end{document}